\documentclass{czjphys}         
\begin{document}
\title{$\mathcal{PT}$-Symmetric Quantum Field Theory}
\authori{Kimball A. Milton}
\addressi{Department of Physics and Astronomy, University of Oklahoma,
Norman, OK 73019 USA}
\authorii{}     \addressii{}
\authoriii{}    \addressiii{}
\authoriv{}     \addressiv{}
\authorv{}      \addressv{}
\authorvi{}     \addressvi{}
\headauthor{K. A. Milton}
\headtitle{$\mathcal{PT}$-Symmetric Quantum Field Theory}
\lastevenhead{K. A. Milton: $\mathcal{PT}$-Symmetric Quantum Field Theory}
\refnum{A}
\daterec{XXX}    
\issuenumber{0}  \year{2003}
\setcounter{page}{1}

\maketitle
\begin{abstract}
In the context of the $\mathcal{PT}$-symmetric version of quantum
electrodynamics, it is argued that the $\mathcal{C}$ operator introduced
in order to define a unitary inner product has nothing to do with
charge conjugation.
\end{abstract}

\section{Introduction}     

A new approach to quantum theories was proposed in 1996 in the context of
the so-called $\delta$-expansion \cite{bm97}.
The idea was to study non-Hermitian
theories which nevertheless possessed positive spectra. Parity 
symmetry $\mathcal{P}$ was broken in these theories, as well as 
time-reversal invariance
$\mathcal{T}$. However, $\mathcal{PT}$ symmetry was unbroken.
 In early papers we examined scalar field theories with
interaction \cite{bm97}
\begin{equation}
\mathcal{L}_{\rm int}=-g({\rm i}\phi)^N,
\end{equation}
supersymmetric theories possessing the superpotential \cite{bm98}
\begin{equation}
\mathcal{S}=-{\rm i}g({\rm i}\phi)^N,\end{equation}
and massless electrodynamics with an axial vector current \cite{bm99}
\begin{equation}
j_5^\mu=e\frac12\psi\gamma^0\gamma^\mu\gamma^5\psi,
\end{equation}
as well as studied the Schwinger-Dyson equations for such theories 
as \cite{bms00}
\begin{equation}
{\cal L}_{\rm int}=-g\phi^4.
\end{equation}

Many remarkable features were discovered in such investigations, such
as  perturbative parity violation, stability of eigenvalue conditions
for coupling constants,
and asymptotic freedom. The idea is essentially
nonperturbative in concept, because the path integrals defining the theory,
in general, must be defined as nontrivial contours in the complex plane.

However, most of what is known about $\mathcal{PT}$-symmetric theories comes
from examples in quantum mechanics, that is, $d=1$ quantum field theory.
For example, it has been proved that the spectrum of
\begin{equation}
H=p^2+x^2({\rm i}x)^\nu,\quad\nu\ge0,
\label{ham}
\end{equation}
is real and positive \cite{bbetc}.

\section{Unitarity}

The most troubling aspect of $\mathcal{PT}$-symmetric theories has been that of
the apparent violation of unitarity.  In a remarkable development last
year it was discovered how to define a unitary norm, at least for 
quantum mechanical theories like that described by the Hamiltonian (\ref{ham}).
Let $\phi_n$ by the $n$th eigenfunction of $H$,
\begin{equation}
H\phi_n(x)=E_n\phi_n(x),
\end{equation}
which is a differential equation imposed on a complex contour $C$.  The 
eigenfunctions
 can be chosen to have eigenvalue 1 of the $\mathcal{PT}$ operator:
\begin{equation}
\mathcal{PT}\phi_n(x)=\phi_n(x).
\end{equation}
The eigenfunctions are complete in the sense that
\begin{equation}
\sum_n(-1)^n\phi_n(x)\phi_n(y)=\delta(x-y),
\end{equation}
which means that the $\mathcal{PT}$ inner product,
\begin{equation}
(f,g)\equiv\int_C {\rm d}x\,
[\mathcal{PT}f(x)]g(x),\quad \mathcal{PT}f(x)=f^*(-x),
\end{equation}
(the complex conjugation is to be applied to quantities in the functional
form, not to the coordinate $x$)
defines a metric which is not definite:
\begin{equation}
(\phi_n,\phi_m)=(-1)^n\delta_{mn}.
\end{equation}

Because of the severe interpretational issues associated with an indefinite
metric, it is fortunate that last year Bender, Brody, and Jones \cite{bbj}
discovered how to define a positive metric.  In terms of the
eigenfunctions, they defined a (dynamical) $\mathcal{C}$ operator:
\begin{equation}
\mathcal{C}(x,y)=\sum_n\phi_n(x)\phi_n(y),
\end{equation}
which has square unity:
\begin{equation}
\int {\rm d}y\,\mathcal{C}(x,y)\mathcal{C}(y,z)=\delta(x-z),
\end{equation}
but is distinct from the parity operator:
\begin{equation}
\mathcal{P}(x,y)=\delta(x+y)=\sum_n(-1)^n\phi_n(x)\phi_n(-y),
\end{equation}
The positive-definite inner product is now defined by
\begin{equation}
\langle f|g\rangle=\int_C{\rm d}x\,[\mathcal{CPT}f(x)]g(x).
\end{equation}
This defines a nontrivial extension of quantum mechanics.
Of course, the physical interpretation of $\mathcal{C}$ is far from clear.

\section{Electrodynamics}
The notation suggests that $\mathcal{C}$ is some sort of charge-conjugation
operator.  The place to examine such an idea would seem to be the fermion
sector, since it was from the Dirac equation that the concept of antiparticles
emerged.  Consider the massless 
Dirac Lagrangian, written in term of the Majorana representation:
\begin{equation}
\mathcal{L}=-\frac12\psi\gamma^0\gamma^\mu\frac1{\rm i}\partial_\mu\psi.
\end{equation}
The symmetry of $\gamma^0\gamma^\mu$, combined with the antisymmetry of
the derivative operator, requires that the Dirac field $\psi$ be a Grassmann
variable.

There are thus two ways to introduce interactions. In either one, one
starts from a global transformation that leaves the Dirac Lagrangian invariant:
\begin{equation}
\psi\to {\rm e}^{{\rm i}\theta\lambda}\psi,
\end{equation}
where $\lambda$ is a constant, and $\theta$ is an antisymmetric  matrix
that commutes with $\gamma^0\gamma^\mu$.

\begin{itemize}
\item The first choice is $\theta=eq$,
\begin{equation}
q=\left(\begin{array}{cc}
0&-{\rm i}\\
{\rm i}&0\end{array}\right),
\end{equation}
the antisymmetrical charge matrix, living in an independent two-dimensional
space.  When $\lambda$ is now promoted to a space-time dependent function,
we are led to the usual current of electrodynamics:
\begin{equation}
j^\mu=\frac12\psi\gamma^0\gamma^\mu eq\psi,
\end{equation}
because the gauge transformation of $A_\mu$, $A_\mu\to A_\mu+\partial_\mu
\lambda$ in the interaction $j^\mu A_\mu$
 cancels that of the free Lagrangian.
It is the additional two-fold multiplicity of the Dirac
 field that corresponds to the presence of antiparticles.
 
 Schwinger sharpened this argument.  He insisted on the 
Euclidean postulate---that the extension of the theory to Euclidean
space bear no memory of the original timelike direction---and showed
therefore that every spin 1/2 particle must have a chargelike
attribute \cite{js70}.

\item The second possibility is that $\theta=e\gamma^5$; this leads
to the axial-current interaction $j_5^\mu A_\mu$, with
\begin{equation}
j_5^\mu=\frac12\psi\gamma^0\gamma^\mu e\gamma^5\psi.
\end{equation}
For a massless theory, such a current is conserved, barring anomalies,
so we still have a consistent theory. 
This is the $\mathcal{PT}$-symmetric QED referred to above.
 However, there is now no additional
two-fold multiplicity of the Dirac field, and therefore, apparently,
no antiparticles. We therefore suspect that referring to the dynamical
$\mathcal{C}$ operator as charge conjugation may be misleading.
\end{itemize}

\section{Discussion}
It is not yet clear if there exists a consistent treatment of fermions
in the $\mathcal{PT}$-symmetric framework.
The exploration of the Dirac
equation, both at the classical and second-quantized level, promises to shed
light on the connection of charge-conjugation, the $\mathcal{C}$ operator,
and unitarity in this exciting extension of quantum mechanical ideas.
I hope that this summary will provoke substantial developments.  Further
developments will be indicated in my contribution to the Workshop on 
Pseudo-Hermitian Hamiltonians.

\bigskip
{\small I thank the US Department of Energy for partial support of
this research, and Qinghai Wang for extensive discussions at this
Colloquium.  I am grateful to the Organizers, particularly Cestmir Burdik,
 of this Colloquium for allowing me to present my work here.}


\bigskip

\bbib{9}               
\bibitem{bm97} C. M. Bender and K. A. Milton, 
Phys.\ Rev.\ D {\bf 55}  (1997) R3255.
\bibitem{bm98} C. M. Bender and K. A. Milton, 
Phys.\ Rev.\ D {\bf 57} (1998) 3595.
\bibitem{bm99} C. M. Bender and K. A. Milton, J. Phys.\ A  {\bf 32 }(1999) L87.
\bibitem{bms00} C. M. Bender, K. A. Milton, and V. M. Savage, Phys.\ Rev.\
D {\bf 62} (2000) 085001.
\bibitem{bbetc} C. M. Bender and S. Boettcher, Phys.\ Rev.\
Lett.\ {\bf 80} (1998) 5243; P. Dorey, C. Dunning, and R. Tateo, J. Phys.\
A {\bf 34} (2001) L391; {\bf 34} (2001)  5679.
\bibitem{bbj} C. M. Bender, D. C. Brody, and H. F. Jones, Phys.\ Rev.\ Lett. 
{\bf 89} (2002) 270401.
\bibitem{js70} J. Schwinger, {\it Particles, Sources, and
Fields\/} (Addison-Wesley, Reading, 1970).
\ebib
\end{document}